\documentclass[fleqn]{2023SCGE}
\usepackage{lastpage}
\usepackage{graphicx}
\usepackage{subfigure}

\begin{document}
\ArticleType{Article}
\SpecialTopic{SPECIAL TOPIC: }

\title{Lunar and Terrestrial Time Transformation Based on the  Principle of General Relativity}

\author[1]{Min Liu}{}
\author[2]{Jing-Song Ping}{}
\author[2]{Wen-Xiao Li}{}
\author[3]{Zhou-Jian Cao}{}
\author[4]{Jie Yang}{}
\author[5]{Yong-Jun Wang}{}
\author[2]{\\Hong-Bo Jin}{}
\author[2,3]{Wen-Zhao Zhang}{}
\author[2,6]{Ming-Xue Shao}{}
\author[7]{Jian-Guo Yan}{}
\author[1]{He-Zhen Yu}{}
\AuthorMark{Min Liu}
\AuthorCitation{Liu M, Ping J S, Li W X, et al.}

\address[1]{Beijing Orient Institute of Metrology and Test, Beijing 100094, China}
\address[2]{National Astronomical Observatories, Chinese Academy of Sciences, Beijing 100101, China}
\address[3]{Department of Physics and Astronomic, Beijing Normal University, Beijing 100091, China }
\address[4]{Lanzhou Center for Theoretical Physics, Key Laboratory of Theoretical Physics of Gansu Province, \\
and Key Laboratory of Quantum Theory and Applications of MoE, Lanzhou University, Lanzhou, Gansu 730000, China}
\address[5]{Lanzhou institute of physics, Lanzhou 73000, China}
\address[6]{Institute of Electrical Engineering of the Chinese Academy of Sciences, Beijing 100190, China}
\address[7]{State Key Laboratory  of Information  Engineering  in  Surveying, Mapping  and Remote  Sensing, Wuhan  University,  Wuhan 430070,China}
\date{\today}
\abstract{Lunar time metrology exemplifies the need for a unified temporal framework beyond Earth. Similar to Earth, the Moon requires an independent system for timekeeping, time dissemination, and calendrical reckoning. Two recent publications by American scholars define and describe lunar time within the relativistic framework, proposing a formula for the transformation between Terrestrial Time (TT) and Lunar Coordinate Time (LTC). However, the derivation and underlying assumptions of this formula have been questioned.
The complex rotational dynamics among massive bodies in the solar system can be simplified by decomposing them into hierarchical wide-area and local-area relationships using the concepts of “external” and “internal” problems. Grounded in the symmetry and conservation principles of physical laws, Einstein’s general relativity emphasizes two key ideas: (i) equal weighting, which posits that relationships among multi-level coordinate systems are independent and self-similar, analogous to fractal geometry; and (ii) locality, which requires that the laws of physics retain invariant forms only in local coordinate systems. Specifically, a non-rotating coordinate system corresponds to the Frenet frame along a particle’s geodesic trajectory. Only by restricting rotating references to the local domain — rather than invoking distant celestial-bodies — can the invariance of physical laws be preserved. The use of distant celestial-bodies to define the orientation of the Earth-centered coordinate system violates the principle of general relativity.
Here, we derive the relationship between coordinate time and proper time, employing the Earth-Moon coordinate system as an intermediary to obtain a simplified transformation formula between LTC and TT. An independent and universal lunar standard time framework is proposed. Notably, the coordinate time transformation coefficient exhibits long-term secular variation, which can be measured and predicted via Earth-Moon time comparisons.}


\maketitle

\section{Introduction}
Establishing a rigorous transformation between coordinate times defined in different reference systems is a foundational issue in general relativity and a newly emerging challenge in time metrology. In 2021, the China Association for Science and Technology released the strategic question “Is there a unified time rule beyond Earth?”. The accompanying position paper argues that Terrestrial Time (TT) and its dissemination signals are valid only in the vicinity of Earth, and that simultaneity is intrinsically confined to a local coordinate system. Consequently, proper times measured in two distinct barycentric frames cannot be synchronised directly; instead, a dedicated procedure is required to unify the temporal scales of separate local systems. Whether the Moon, Mars, the Galilean satellites, or the moons of Saturn should possess independent timing systems has thus become a central topic in international astronomy and metrology\cite{liu2022discussion}.
The Moon should indeed be equipped with an autonomous timing infrastructure whose timekeeping, time-transfer, and calendrical rules differ fundamentally from those on Earth. On 2 April 2024, the White House Office of Science and Technology Policy (OSTP) issued the memorandum “Celestial Time Standardization Policy”\cite{whitehouse2024celestial}, directing NASA to develop a unified standard for the Moon and other celestial bodies, with an initial focus on cislunar and lunar-surface missions. Coordinated Lunar Time (LTC) is expected to be defined by the end of 2026. Owing to the Moon’s weaker gravitational potential, LTC will advance with respect to TT by approximately 58.7$\mu s day^{-1}$; additional periodic terms will further increase the divergence.
The International Astronomical Union (IAU) Resolution B1.3 (2000) provides the transformation between Barycentric Coordinate Time (TCB) of the Solar-System barycentric reference system (BCRS) and Geocentric Coordinate Time (TCG) of the Geocentric Celestial Reference System (GCRS). Resolution B1.5 (2000) refines the TCB–TCG relation by incorporating post-Newtonian potential coefficients\cite{kaplan2005iau}. Authors extend these formulae to the Lunar Celestial Reference System (LCRS) in the paper\cite{kopeikin2024lunar}. To relate Lunar Coordinate Time (TCL) to TCG, they invoke TCB as an intermediate variable and obtain the TCB – TCL relation by a simple relabelling of indices. This derivation (i) is not sensitive to the gravitational constraints imposed by the Schwarzschild metric and (ii) merely re-expresses terrestrial time on the Moon, thereby failing to provide a physically independent definition of lunar standard time. 
\begin{figure*}
\includegraphics[width=0.9\textwidth]{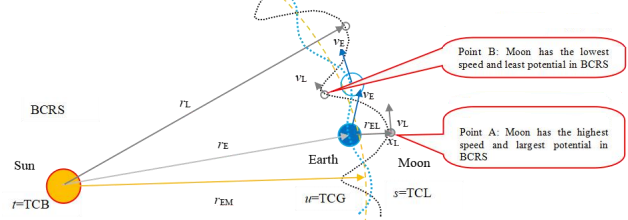}
\caption{In the solar system, the Earth and the Moon orbit around the barycenter they form, resulting in periodic changes in the gravitational potential exerted by the Sun.}
 \label{fig1}
\end{figure*}
 \section{Schwarzschild-metric prerequisites}  
When Schwarzschild first integrated the Einstein field equations he imposed four explicit boundary conditions on the space-time metric \cite{zhao2010fundamentals}: (i) spherical symmetry, (ii) staticity, (iii) a mass distribution confined to a finite spatial region, and (iv) an origin located at the system barycentre. Any observer whose coordinate origin does not coincide with this barycentre—and who therefore retains the influence of additional masses—violates at least one of the above constraints (symmetry and boundedness in particular).
In the paper\cite{kopeikin2024lunar}, there is a purely formal procedure: the subscript ‘E’ (Earth) in the IAU 2000 Resolution B1.3 expression is replaced by ‘L’ (Lunar), while equation (1) is referred verbatim from the Schwarzschild metric underlying that resolution. Two puzzlements follow directly:
\begin{itemize}
\item The IAU TCB–TCG transformation\cite{soffel2003iau} is strictly valid for the Earth. Its recommended metric is tacitly required to satisfy the Schwarzschild conditions. The lunar barycentre, however, is neither spherically symmetric with respect to the BCRS origin nor embedded in a static space-time; the terrestrial gravitational perturbation dominates. Within the BCRS the Moon follows a wavelike trajectory whose speed varies between 28.7 $km s^{-1}$ and 30.6 $km s^{-1}$, implying a 12\%  relative change in kinetic energy. At position A in Picture \ref{fig1} the Moon has maximum potential and kinetic energy; at position B both are minimal. Inside the integral the mechanical energy is not conserved and the Hamiltonian exhibits a time-dependent modulation; hence the required symmetries are broken. Consequently, a mere relabelling of the subscript from the expression (1) to (2) in the paper\cite{kopeikin2024lunar} deviates the Schwarzschild prerequisites of spherical symmetry, statically and bounded mass distribution.

\item Einstein’s general relativity is conventionally formulated under the torsion-free condition (Cartan torsion = 0). If the Earth–Moon barycentre is treated as a single point particle within the BCRS, or if the Earth alone is considered owing to its dominant mass, this condition can be met approximately. Transitioning from the BCRS to the Lunar Celestial Reference System (LCRS), however, necessarily traverses the Earth–Moon-Centre Reference System (EMCRS). The three-dimensional trajectory of the lunar barycentre expressed in BCRS coordinates x$_L$ is therefore endowed with non-vanishing torsion. E is replaced by L index substitution ignoring this geometric constraint; the derivation is valid only as a post-Newtonian approximation whose torsion terms have been truncated.
 \end{itemize}
 
 \section{The Wide-Area ``External Problem'' and the Local ``Internal Problem''}

In the two-body problem, celestial motions obey Keplerian orbital conditions; consequently, the physical laws within the local coordinate frame must exhibit both symmetry and conservation.  
Taking these symmetries and conservation laws as the starting point, we reduce the intricate mutual rotations among celestial bodies to a hierarchy of nested wide-area and local-area relationships, each level being approximated by a two-body kinematic model.

Wide area (external problem): If the collective behaviour of a group of bodies in the upper-level frame is represented as a single point mass, that frame is termed the \emph{wide area}.  
The free-fall motion of this composite mass is called the \emph{external problem} \cite{damour1991general}.

Local area (internal problem):
  A coordinate system whose origin is fixed at the barycentre of the same group constitutes the lower-level frame, referred to as the \emph{local area}.  
  The individual motions of the bodies inside this frame are called the \emph{internal problem}.

In 1991, the paper \cite{damour1991general} has showed that when solving the internal problem, the gravitational potential of the external wide area—including the so-called tidal potential of external bodies—can be entirely neglected.

Post-Newtonian celestial mechanics, however, does not introduce the concept of multi-level nested frames; it operates within a single spatial reference system and does not distinguish between the external and internal problems.  
Consequently, it retains tidal-potential corrections, and the resulting formulae fail to preserve local conservation laws and symmetries.

Ashby \& Patla \cite{ashby2024relativistic} derive a transformation between proper time on the lunar surface ($\tau_{\mathrm{LT}}$) and proper time on the terrestrial geoid ($\tau_{\mathrm{TT}}$) by imposing free-fall and Keplerian conditions.  
They adopt the true anomaly of the elliptic orbit as the independent variable instead of coordinate time, skilfully circumventing the irregularity of the time integral.  
Nevertheless, the Earth--Moon-centre reference system (EMCRS) employed by Ashby \& Patla does not demarcate the boundary between the internal and external problems; the solar potential exterior to the EMCRS is still treated as a tidal term.  
Their derivation is therefore an \emph{ad-hoc} recipe that merely adapts terrestrial standard time to the lunar environment, rather than defining an autonomous lunar standard time.  
The absence of a multi-level nesting within the EMCRS and the lack of a rigorous recursive procedure are likely the principal reasons why Ref.~\cite{ashby2024relativistic} conflates the internal and external problems.

Reducing the many-body problem to multi-level nested coordinate systems, each containing only a two-body problem, is conditional: the characteristic size of the local domain must be far smaller than the curvature radius of the wide area.  
For the Earth--Moon system orbiting within the Solar System, the curvature radius is approximately $1.49\times10^{11}\,\mathrm{m}$, whereas the semi-major axis of the Earth--Moon barycentric orbit is $7.69\times10^{8}\,\mathrm{m}$ — a ratio of about 194.  
This separation of scales justifies treating the Sun–Earth–Moon three-body problem as two nested two-body systems: (i) Solar-System--EMCRS and (ii) Earth--Moon.
\section{Rotation of the Local Coordinate Frame}
\begin{figure*}
\includegraphics[width=0.9\textwidth]{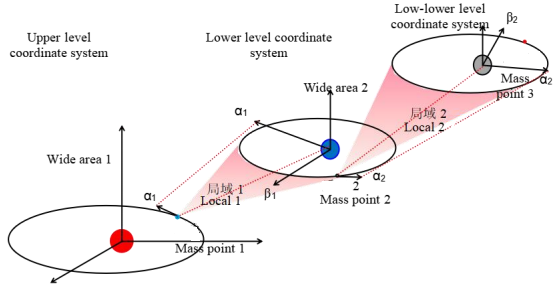}
\caption{The idea of equal rights in general relativity principle and self-similarity in fractal geometry.}
 \label{fig2}
\end{figure*}
Post-Newtonian celestial mechanics designates the Barycentric Celestial Reference System (BCRS) as an inertial frame by stipulating that the BCRS does not rotate with respect to distant extragalactic objects.  
GCRS is then declared quasi-inertial, i.e.\ non-rotating relative to the same remote sources.  
Invoking a reference direction that straddles two hierarchical levels complicates the coordinate-time transformation formulas and, more importantly, imports the \emph{external problem} into the \emph{internal problem}, thereby breaking local conservation laws and symmetries.  
Anchoring the orientation of the GCRS to remote celestial-bodies violates the principle of general relativity.  
Relinquishing the requirement that the GCRS be non-rotating with respect to those celestial-bodies and, instead, selecting a rotationally local reference within the next-higher (but still local) level restores compliance with general relativity and yields a far simpler time-transformation law.

The principle of general relativity embodies two complementary ideas: \emph{equal footing} and \emph{locality}.  
Equal footing asserts that the relationship between upper- and lower-level frames is one of equal standing: physical laws replicate themselves from level to level with the same functional form, much like the self-similarity of fractal geometry \cite{li2000fractal}.  
In the upper-level frame, the intricate internal structure of a lower-level system is ``hidden'' inside a single point particle whose motion is assumed to be non-rotational; no internal details are retained in the upper level.  
This hierarchical decoupling is illustrated in Picture \ref{fig2}.

In Picture \ref{fig2}, particle~1 executes a Keplerian orbit around its central body O$_{1}$.  
Within the wide-area frame (level~1) this motion enjoys Keplerian symmetry and Hamiltonian conservation.  
Embedded in particle~1 is the local frame (level~1) whose spatial orientation $\alpha_{1}$ is aligned with the tangent to the geodesic (i.e.\ the velocity vector) of particle~1 in the wide-area frame.  
Particle~2 orbits its own central body O$_{2}$; again, symmetry and conservation hold within its local frame.  
Recursively, particle~3 and O$_{3}$ reside inside particle~2, and the same symmetries apply in the even deeper local frame.  

However, if one attempts to describe the motion of particle~3 directly in the outermost wide-area frame—e.g.\ by expressing the angular velocity of particle~3 about O$_{3}$ with respect to O$_{1}$—one must superimpose the orbit of particle~2 within level~1 and that of particle~1 within the wide-area frame.  
Such cross-level mixing destroys the symmetries and conservation laws that are intrinsic to the local frame of particle~3.

\begin{figure*}
\includegraphics[width=0.9\textwidth]{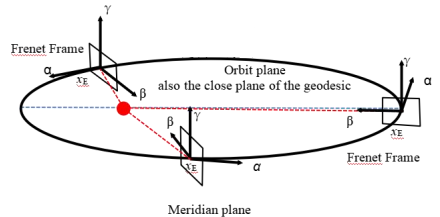}
\caption{Schematic representation of the Ferret frame of the curve.}
 \label{fig3}
\end{figure*}

The principle of locality holds that physical laws — such as the symmetries and conservation laws embodied by Noether’s theorem — are valid only within a local coordinate system \cite{cameron2012electric}.  
Starting from a locally defined natural reference, the non-rotating frame attached to a freely falling particle is the \emph{Frenet frame} $\{\vec\alpha,\vec\beta,\vec\gamma\}$ of differential geometry \cite{li2000fractal}.  
Picture \ref{fig3} illustrates this construction, where  
\[
\vec\alpha=\text{unit tangent to the world-line}, \quad  
\vec\gamma=\text{principal normal}, \quad 
\vec\beta=\vec\alpha\times\vec\gamma=\text{binormal}.
\]  
In general relativity, the barycentric frame carried by a freely falling test particle is often referred to as a Fermi coordinate system \cite{han1990geocentric}.  
While both the Frenet and Fermi frames move along with the particle, they differ in emphasis:  
the Frenet frame rigidly aligns one spatial axis with the instantaneous velocity vector, whereas the Fermi frame enforces geodesic motion but does \emph{not} prescribe the orientation of its spatial triad.  
We therefore adopt the Frenet frame to characterise the rotational properties of free-fall motion.

The orbital tangent $\vec\alpha$ coincides with the three-dimensional velocity and changes continuously along the trajectory, so a fixed forward reference cannot be attached to it.  
To obtain a stable reference against the stellar background, the $\vec\beta\text{--}\vec\gamma$ plane is chosen as the \emph{meridian plane}.  
For a circular orbit, $\vec\beta$ points to the geometric centre, which, in the BCRS, is approximately the solar position.  
For an elliptical orbit, $\vec\beta$ points to the occupied focus only at the apsides; at other times it deviates slightly but remains close to the BCRS origin.  
The focus therefore oscillates periodically across the meridian plane — sometimes leading, sometimes lagging — preserving global periodicity over one revolution.  
The familiar equation-of-time difference between apparent and mean solar days observed from Earth is a direct consequence of this mechanism \cite{han2018principle}.  
Although the instantaneous angular velocity of the Frenet frame with respect to the focal reference is non-uniform, the cycle-averaged value is well-defined and can be employed to determine the exact start and end points of each orbital period.

\section{Proper-time to Coordinate-time Conversion (\(\tau\leftrightarrow t\))}
From the locality principle of general relativity, a time transformation must be carried out \emph{within one and only one coordinate system}; it is illegitimate to convert coordinate times that belong to \emph{different hierarchical levels}.  
The IAU TCB–TCG relation in fact straddles two levels, so the two quantities should not appear together in a single conversion formula.  
Because space and time are inseparable, \emph{every time scale must be tied to a definite spatial coordinate}, and the latter must be expressed by a \emph{stable, periodic ephemeris}.

Hence the only legitimate transformation is \emph{coordinate time \(\leftrightarrow\) proper time} inside one frame; a transformation labelled “coordinate time \(\leftrightarrow\) coordinate time’’ is ill-defined and leads to ambiguities.  
The IAU TCB–TCG formula may be re-interpreted as follows: TCG is the proper time at the Earth-barycentre position \(\vec x_{\!E}\) \emph{within} the BCRS, while the notional observer’s clock is rigidly affixed to the Earth barycentre and \emph{cannot be located elsewhere}.

\subsection*{5.1\quad \(\tau\)–\(t\) relation inside one frame}

For a metric \(g_{\mu\nu}\) in a given coordinate system the line element gives
\begin{equation}
\mathrm d\tau^{2}=-\frac{1}{c^{2}}\,g_{\mu\nu}\,\mathrm dx^{\mu}\mathrm dx^{\nu}.
\end{equation}

Setting spatial displacements to zero (\(\mathrm d\vec x=0\)) and expanding to 1\,PN accuracy, one obtains
\begin{equation}
\Delta t=\int_{\tau_{0}}^{\tau_{1}}\!\Bigl(1+\frac{w}{c^{2}}+\frac{1}{2}\frac{v^{2}}{c^{2}}\Bigr)\mathrm d\tau.
\label{eq:tau-t}
\end{equation}
Here
 \(t\)  coordinate time,
   \(\tau\)  proper time,
   \(w\)  gravitational potential at the observer,
   \(\vec v\)  velocity relative to the coordinate origin,
  \(c\)  speed of light.

In the Solar System the solar surface potential amounts to only \(2.1\times10^{-6}\); the weak-field (1\,PN) approximation therefore suffices, and frame–dragging is negligible for the Earth–Moon system \cite{ni2016solar}.

Equation \eqref{eq:tau-t} yields \emph{coordinate time from measured proper time}; the bracketed terms reflect orbital potential and kinetic energies.  
For a freely falling test particle moving along a geodesic, the sum of kinetic and potential energy is conserved, guaranteeing the \emph{time-translation invariance} of the Hamiltonian \(H\) over one orbital period \(T\):
\begin{equation}
H\,T=\text{const.}
\end{equation}

Note that the velocity \(\vec v\) in \eqref{eq:tau-t} is expressed in the \emph{Frenet frame} (\(\alpha\)--EMCRS), whose axes are aligned with the geodesic tangent \(\vec\alpha\).  
After one heliocentric revolution, \(\vec\alpha\) rotates through \(360^{\circ}\) relative to the BCRS fixed on distant quasars; the resulting difference in \(v^{2}\) absorbs the solar tidal potential within the \(\alpha\)--EMCRS — an immediate consequence of the \emph{locality} requirement.

\subsection*{5.2\quad Hierarchical nesting: TCB\(\rightarrow\)TCC}

Whenever the integrand in \eqref{eq:tau-t} is constant (conserved Hamiltonian and free-fall motion), the coefficient can be taken outside the integral:
\begin{equation}
\Delta t=\Bigl(1+\frac{w}{c^{2}}-\frac{1}{2}\frac{v^{2}}{c^{2}}\Bigr)\Delta\tau.
\label{eq:linear-conv}
\end{equation}

The IAU 2000 formulation omits the intermediate \emph{Earth–Moon barycentric frame} (EMCRS) by letting the Earth barycentre orbit the BCRS directly; lunar effects are then treated as tidal perturbations, producing cumbersome series.

By contrast, the \emph{\(\alpha\)--EMCRS} moves on an equipotential surface of the Sun; external potentials are relegated to the \emph{external problem} and generate \emph{no tides} inside the frame.  
General relativity asserts the \emph{equal standing} of all frames: TCB relates to the coordinate time \emph{TCC} of the Earth–Moon barycentre (OEM) via \eqref{eq:linear-conv} \emph{inside the BCRS}, whereas TCC and TCG are linked \emph{inside the \(\alpha\)--EMCRS}.  
A direct TCB\(\leftrightarrow\)TCG formula therefore straddles levels and violates hierarchical symmetry.

Assuming periodic OEM ephemerides and negligible mass loss, \eqref{eq:linear-conv} yields
\begin{equation}
\Delta\mathrm{TCC}=\Bigl(1+\frac{w_{\mathrm C}}{c^{2}}-\frac{1}{2}\frac{v_{\mathrm C}^{2}}{c^{2}}\Bigr)\Delta\mathrm{TCB},
\label{eq:TCC-TCB}
\end{equation}
with
 \(w_{\mathrm C}\), \(\vec v_{\mathrm C}\)  gravitational potential and velocity of OEM in the BCRS,
 \(L_{\mathrm{CC}}\)  dimensionless conversion coefficient (slowly varying).

\subsection*{5.3\quad TCG\(\leftrightarrow\)TCL via \(\alpha\)--EMCRS}

Inside the \(\alpha\)--EMCRS, Earth and the Moon describe coplanar ellipses about their common barycentre OEM (\(\approx4\,600\) km from Earth’s centre).  
Applying \eqref{eq:linear-conv} to each body gives
\begin{equation}
\Delta\mathrm{TCG}=\Bigl(1+\frac{w_{\mathrm{CE}}}{c^{2}}-\frac{1}{2}\frac{v_{\mathrm{CE}}^{2}}{c^{2}}\Bigr)\Delta\mathrm{TCC},
\label{eq:TCG-TCC}
\end{equation}

\begin{equation}
\Delta\mathrm{TCL}=\Bigl(1+\frac{w_{\mathrm{CL}}}{c^{2}}-\frac{1}{2}\frac{v_{\mathrm{CL}}^{2}}{c^{2}}\Bigr)\Delta\mathrm{TCC}.
\label{eq:TCL-TCC}
\end{equation}
Here the subscripts CE and CL denote quantities at the Earth centre and the lunar centre, respectively, within the \(\alpha\)--EMCRS.

Eliminating TCC from \eqref{eq:TCG-TCC} and \eqref{eq:TCL-TCC} yields
\begin{equation}
\Delta\mathrm{TCL}=\bigl(1+L_{\mathrm{GL}}\bigr)\,\Delta\mathrm{TCG},
\qquad
L_{\mathrm{GL}}\approx\frac{w_{\mathrm{CL}}-w_{\mathrm{CE}}}{c^{2}}+\frac{v_{\mathrm{CE}}^{2}-v_{\mathrm{CL}}^{2}}{2c^{2}}.
\label{eq:TCL-TCG}
\end{equation}
Because both bodies are in free fall and their ephemerides are periodic, \(L_{\mathrm{GL}}\) is a \emph{quasi-constant} whose long-term drift reflects slow mass- and energy-exchange with external bodies.  
Rather than measuring the orbital parameters directly, one may
\begin{itemize}
  \item realise SI seconds on Earth and on the Moon with independent clocks,
  \item compare proper times via Earth–Moon links,
  \item iteratively refine \(L_{\mathrm{GL}}\).
\end{itemize}
This procedure respects the \emph{equal standing} of the GCRS and the LCRS and avoids the proliferation of tidal perturbation series.
\section{Hierarchically Nested Coordinate Structure (TCB–TCC Relation)}

When the integrand in \eqref{eq:linear-conv} is strictly constant — i.e.\ the system’s mechanical energy is conserved and the barycentre is in free fall — the coefficient can be taken outside the integral, yielding a \emph{linear} relation between coordinate time and proper time:

\begin{equation}
\Delta t = \Bigl(1+\frac{w}{c^{2}}-\frac{1}{2}\frac{v^{2}}{c^{2}}\Bigr)\Delta\tau \equiv (1+L)\Delta\tau.
\label{eq:linear-L}
\end{equation}

The conventional IAU 2000 TCB–TCG formulation \emph{compresses the hierarchy} by omitting the intermediate \emph{Earth–Moon barycentric coordinate system} (EMCRS).  It approximates the Earth barycentre as a test particle orbiting the BCRS, with known orbital parameters $(a_E,\,e_E,\,i_E,\,\Omega_E,\,\omega_E,\,M_E(t))$, and treats lunar effects via tidal perturbations.  
This two-level jump inevitably leads to \emph{non-uniform} TCB–TCG relations and necessitates intricate perturbation series.

In contrast, the \emph{$\alpha$-EMCRS} moves on an equipotential surface of the Solar gravitational field.  External potentials (Sun + planets) are relegated to the \emph{external problem} and, by construction, exert \emph{no tidal forces} on the interior of $\alpha$-EMCRS.  

General relativity dictates that \emph{all coordinate systems are on an equal footing}.  
Hence
\begin{itemize}
  \item TCB relates to the coordinate time \emph{TCC} of the Earth–Moon barycentre (OEM) via \eqref{eq:linear-L} \emph{inside the BCRS};
  \item TCC and TCG are linked by \eqref{eq:linear-L} \emph{inside the $\alpha$-EMCRS}.
\end{itemize}
A direct TCB$\leftrightarrow$TCG formula \emph{violates the layering} and mixes hierarchical levels.

Adopting the OEM ephemeris $(x_C,v_C)$ as periodic and neglecting slow mass loss, one obtains

\begin{equation}
\Delta\mathrm{TCC} = \Bigl(1+\frac{w_C}{c^{2}}-\frac{1}{2}\frac{v_C^{2}}{c^{2}}\Bigr)\Delta\mathrm{TCB} \equiv (1+L_{\mathrm{CC}})\Delta\mathrm{TCB},
\label{eq:TCC-TCB-L}
\end{equation}

where
 $w_C$ gravitational potential at OEM in the BCRS,
 $v_C$  speed of OEM relative to BCRS,
 $L_{\mathrm{CC}}$ dimensionless conversion coefficient \emph{for the BCRS level}.

The hierarchical structure is therefore

\[
\text{TCB} \xrightarrow{L_{\mathrm{CC}}} \text{TCC} \xrightarrow{L_{\mathrm{CG}}} \text{TCG}.
\]

This replaces the single IAU 2000 formula by two physically transparent steps, each respecting locality and conservation.

\section{Earth--Moon Relationship in the Synodic $\alpha$-EMCRS Frame (TCL$\leftrightarrow$TCG)}

Inside the $\alpha$\,--Earth–Moon barycentric reference system ($\alpha$\,--EMCRS) both Earth and the Moon describe coplanar, elliptical two-body orbits aabout their common barycentre OEM ($\approx 4\,600$ km from Earth’s centre and inside the Earth).
The coordinate time of this frame is denoted TCC.

Applying the linear conversion \eqref{eq:linear-L} separately at the Earth and Moon centres gives
\begin{align}
\Delta\mathrm{TCG}
    &=\Bigl(1+\frac{w_{\mathrm{CE}}}{c^{2}}
            -\frac{1}{2}\frac{v_{\mathrm{CE}}^{2}}{c^{2}}\Bigr)\Delta\mathrm{TCC}
     \equiv (1+L_{\mathrm{CG}})\Delta\mathrm{TCC},
\label{eq:TCG-TCC-L}\\[2mm]
\Delta\mathrm{TCL}
    &=\Bigl(1+\frac{w_{\mathrm{CL}}}{c^{2}}
            -\frac{1}{2}\frac{v_{\mathrm{CL}}^{2}}{c^{2}}\Bigr)\Delta\mathrm{TCC}
     \equiv (1+L_{\mathrm{CL}})\Delta\mathrm{TCC},
\label{eq:TCL-TCC-L}
\end{align}
where
 $w_{\mathrm{CE}},\;v_{\mathrm{CE}}$  potential and speed of the Earth centre in $\alpha$\,--EMCRS,
$w_{\mathrm{CL}},\;v_{\mathrm{CL}}$  potential and speed of the Moon centre in $\alpha$\,--EMCRS,
 $L_{\mathrm{CG}},\;L_{\mathrm{CL}}$  dimensionless coefficients \emph{inside the $\alpha$\,--EMCRS level}.

Eliminating $\Delta\mathrm{TCC}$ between \eqref{eq:TCG-TCC-L} and \eqref{eq:TCL-TCC-L} yields the direct Earth–Moon relation
\begin{equation}
\Delta\mathrm{TCL}=(1+L_{\mathrm{GL}})\,\Delta\mathrm{TCG},
\qquad
L_{\mathrm{GL}}\approx\frac{w_{\mathrm{CL}}-w_{\mathrm{CE}}}{c^{2}}
                   +\frac{v_{\mathrm{CE}}^{2}-v_{\mathrm{CL}}^{2}}{2c^{2}}.
\label{eq:TCL-TCG-L}
\end{equation}
Because both Earth and the Moon are in free fall within $\alpha$\,--EMCRS and their ephemerides are periodic, $L_{\mathrm{GL}}$ is a \emph{quasi-constant}.  
Slow variations of $L_{\mathrm{GL}}$ arise only from secular mass–energy exchange with the exterior, allowing its long-term behaviour to be measured and predicted.

\paragraph{Operational realisation.}
Instead of determining $(w_{\mathrm{CE}},v_{\mathrm{CE}},w_{\mathrm{CL}},v_{\mathrm{CL}})$ from ephemerides, one
\begin{enumerate}
  \item realises SI seconds independently in atomic clocks on Earth (TCG) and on the Moon (TCL),
  \item exchanges proper time via Earth–Moon communication links,
  \item iteratively refines $L_{\mathrm{GL}}$ through continuous time comparison.
\end{enumerate}
This procedure respects the \emph{equal standing} of the GCRS and the LCRS and avoids complicated tidal perturbation series.

\section{Discussion and Conclusions}

Starting from the relativistic view of time and the principle of general relativity, we have re-examined the coordinate-time transformation prescribed by IAU 2000 Resolution B1.3 (TCB\,--\,TCG) and questioned the validity of the two-level leap adopted in the paper\cite{kopeikin2024lunar}.  
The key points are:

\begin{enumerate}
  \item \textbf{Simultaneity is defined only within a single coordinate system.}  
  Any attempt to convert coordinate times across hierarchical levels violates the \emph{equal--footing} and \emph{locality} tenets of general relativity and breaks the underlying symmetries and conservation laws.

  \item \textbf{A three-level hierarchy is necessary.}  
  By inserting the Earth--Moon barycentric coordinate system ($\alpha$\,--EMCRS) between the BCRS and the individual geocentric/lunar frames, we obtain the \emph{linear} and \emph{transparent} chain
  \[
  \mathrm{TCB}\xrightarrow{L_{\mathrm{CC}}}\mathrm{TCC}
  \xrightarrow{L_{\mathrm{GL}}}\mathrm{TCG},\qquad
  \mathrm{TCB}\xrightarrow{L_{\mathrm{CC}}}\mathrm{TCC}
  \xrightarrow{L_{\mathrm{CL}}}\mathrm{TCL},
  \]
  where each coefficient is derived from \eqref{eq:linear-L} within its own level.

  \item \textbf{Practical prescription for lunar time.}  
  \begin{itemize}
    \item[\textbullet] The lunar surface constitutes an independent barycentric system.  
    A lunar standard time (LST) should be realised as the \emph{coordinate time} TCL of the lunar barycentric frame, anchored to the SI second via on--Moon atomic clocks and an appropriate lunar calendar.

    \item[\textbullet] Earth and the Moon are on an equal footing.  
    After broadcast to the other body, neither TCG nor TCL is regarded as the \emph{standard} for the recipient; instead they serve as \emph{references} for time comparison.  
    The conversion coefficient $L_{\mathrm{GL}}$ is therefore determined \emph{empirically} through continuous two-way time transfer and long-term averaging, rather than by precise orbit determination.

    \item[\textbullet] Operational procedure:  
    (i) realise independent proper-times on Earth and the Moon;  
    (ii) exchange these readings via radio links;  
    (iii) iteratively refine $L_{\mathrm{GL}}$ and issue regular updates of the TCG\,--\,TCL conversion table and a synchronised lunar--terrestrial calendar.
  \end{itemize}

  \item \textbf{Conceptual implications.}  
  The adoption of the Frenet frame as the non-rotating local reference guarantees that the laws of physics retain their invariant form inside each level.  
  Tidal potentials from the exterior are automatically absorbed into the kinetic term $v^{2}/2c^{2}$ and need not be introduced as separate perturbations.

\end{enumerate}

In summary, the proposed framework provides a conceptually consistent, computationally simple, and operationally feasible route to establishing an \emph{independent and universal lunar standard time}.

\section*{References}
\bibliographystyle{scichina}
\bibliography{ref}
\end{document}